\documentclass[12pt]{article}

\usepackage{graphicx}
\usepackage{a4}

\usepackage{fancybox}
\usepackage{epsfig}
\usepackage{color}

\usepackage{amsfonts}
\usepackage{mathrsfs}
\usepackage{amssymb}
\usepackage{amsmath}

\usepackage{dcolumn}
\usepackage{bm}

\def\pa{\partial}

\def\al{\alpha}

\def\de{\delta}
\def\ep{\epsilon}

\def\th{\theta}

\def\si{\sigma}

\def\ps{\psi}
\def\om{\omega}

\def\La{\Lambda}

\def\Om{\Omega}

\newcommand{\ben}{\begin{equation}}
\newcommand{\een}{\end{equation}}
\newcommand{\bea}{\begin{eqnarray}}
\newcommand{\eea}{\end{eqnarray}}
\newcommand{\ba}{\begin{array}}
\newcommand{\ea}{\end{array}}
\newcommand{\bit}{\begin{itemize}}
\newcommand{\eit}{\end{itemize}}

\newcommand{\vs}[1]{\vspace{#1 mm}}

\newcommand{\dsl}{\pa \kern-0.5em /}

\begin{document}

\topmargin 0pt \oddsidemargin 0mm

\vspace{2mm}

\begin{center}

{\Large Black hole magnetospheres in the Born-Infeld theory}

\vs{10}

 {\large Huiquan Li \footnote{E-mail: lhq@ynao.ac.cn} and Jiancheng Wang}

\vspace{6mm}

{\em

Yunnan Observatories, Chinese Academy of Sciences, \\
650216 Kunming, China

Key Laboratory for the Structure and Evolution of Celestial Objects,
\\ Chinese Academy of Sciences, 650216 Kunming, China

Center for Astronomical Mega-Science, Chinese Academy of Sciences,
\\ 100012 Beijing, China}

\end{center}

\vs{9}

\begin{abstract}

We study the force-free electrodynamics on rotating black holes in
the Born-Infeld (BI) effective theory. The stream equation
describing a steady and axisymmetric magnetosphere is derived. From
its near-horizon behavior, we obtain the modified Znajek regularity
condition, with which we find that the horizon resistivity in the BI
theory is generally not a constant. As expected, the outer boundary
condition far away from the hole remains unchanged. In terms of the
conditions at both boundaries, we derive the perturbative solution
of split monopole in the slow rotation limit. It is interesting to
realise that the correction to the solution relies not only on the
parameter in the BI theory, but also on the radius (or the mass) of
the hole. We also show that the quantum effects can undermine the
energy extraction process of the magnetosphere in the non-linear
theory and the extraction rate gets the maximum in the Maxwell
theory.

\end{abstract}



\section{Introduction}
\label{sec:introduction}

Analogous to neutron stars and other ordinary objects,
magnetospheres can also form on astronomical black holes. In the
black hole magnetosphere, plenty of electron-positron pairs can be
created via the pair cascade processes \cite{Blandford:1977ds}. With
strong electromagnetic fields, this form the force-free
magnetosphere in which the electric fields along the magnetic field
lines are screened and the charges feel zero net force.

The force-free magnetosphere can be used to extract the rotational
energy of a rotating black hole. As a kind of Penrose process, the
energy is extracted through the rotation of the magnetosphere
dragged by the black hole spacetime. This process has become a
promising mechanism nowadays that can explain the formation of
powerful jets observed in many high energy objects, like AGN, GRB
and microquasars.

It is known that the magnetic fields on neutron stars are very high,
even exceeding the quantum electrodynamics (QED) critical value. In
this case, the QED corrections should be included in the force-free
magnetospheres and the Maxwell electrodynamics should be replaced by
the non-linear theory. This has been discussed in magnetar
magnetospheres (e.g.,
\cite{Denisov:2003ba,Pereira:2018mnn,Li:2019uws}), whose surface
magnetic fields sometimes can be above $10^{15}$ G.

However, this is not the case for black hole magnetospheres. The
astronomical black holes do not have their own magnetic fields. The
magnetic fields on them come from accretion and are usually far
weaker (e.g., $\lesssim10^4$ G for a black hole with mass $M=10^9$
$M_\odot$ \cite{1978MNRAS.185..833Z}) than the QED critical value.
But, the study of the black hole magnetosphere in non-linear
electrodynamics is useful for that of the QED corrected
magnetosphere near a magnetar where gravity is important. The latter
can be obtained in a weak field limit of the former despite a
difference of the inner boundary condition.

The study of black hole magnetospheres in non-linear electrodynamics
is also of theoretical interest. The force-free magnetospheres on
black holes are not well understood even in the Maxwell theory. We
still do not know well the structure and geometry of the field lines
in the force-free magnetosphere. The extension to the non-linear
theory help find the analytical properties of black hole
magnetospheres in a general sense.

Moreover, non-linear electrodynamics include quantum corrections to
the Maxwell theory. As is known, strong quantum effects also happen
in black holes. Thermal particles are excited and radiated in the
near-horizon regions of black holes. It is interesting to examine
the force-free non-linear electrodynamics in these regions.

In this work, we consider the black hole magnetospheres in
non-linear electrodynamics, in particular in the Born-Infeld (BI)
effective theory \cite{Born:1934gh}. The BI theory has an explicit
expression with well-regularized features. It also arises in string
theory and so attracts much attention. The paper is organized as
follows. In Section \ref{sec:FFnonE}, we present the force-free
theory in general non-linear electrodynamics. In this general
framework, the stream equation describing steady and axisymmetric
magnetospheres on rotating black holes is derived in Section
\ref{sec:BHmagnetosphere}. Based on the stream equation, the
boundary conditions are discussed in Section \ref{sec:bc} and the
perturbative solution of split monopole is derived in Section
\ref{sec:monopolesol}. In the final section, we summarize and
discuss the results.

\section{Force-free non-linear electrodynamics}
\label{sec:FFnonE}

We start with the action of general electrodynamics
\begin{equation}\label{e;EMaction}
 S=\int \sqrt{-g}\left[\frac{1}{4\pi}\mathcal{L}_{\textrm{EM}}(s,p)
+A_\mu J^\mu\right]d^4x,
\end{equation}
where $\mathcal{L}_{\textrm{EM}}(s,p)$ is general Lagrangian of the
electromagnetic fields with
\begin{eqnarray}
 s=\frac{1}{4}F^{\mu\nu}F_{\mu\nu}, &&
p=\frac{1}{4}\widetilde{F}^{\mu\nu}F_{\mu\nu}.
\end{eqnarray}
The dual field strength
$\widetilde{F}^{\mu\nu}=(1/2)\ep^{\mu\nu\rho\si}F_{\rho\si}$. For
the BI theory, the Lagrangian of electromagnetic fields takes the
form
\begin{equation}\label{e:BI}
 \mathcal{L}_{\textrm{EM}}(s,p)=b^2\left(1-\sqrt{1+
\frac{2s}{b^2}-\frac{p^2}{b^4}}\right),
\end{equation}
where $b$ is an undetermined parameter, with the dimension of mass
squared.

The modified Maxwell equations are given by
\begin{equation}
 \nabla_\mu \widetilde{F}^{\mu\nu}=0,
\end{equation}
\begin{equation}\label{e:currents}
 \nabla_\mu G^{\mu\nu}=4\pi J^\nu,
\end{equation}
where $J^\nu$ is the conserved current and
\begin{equation}
 G^{\mu\nu}=SF^{\mu\nu}+P\widetilde{F}^{\mu\nu},
\end{equation}
with
\begin{equation}
 S\equiv\pa_s\mathcal{L}_{\textrm{EM}}, \textrm{ }
\textrm{ }\textrm{ } P\equiv\pa_p\mathcal{L}_{\textrm{EM}}.
\end{equation}

The energy-momentum tensor of the electromagnetic fields is obtained
from the derivative of the Lagrangian with respect to the metric
\begin{equation}\label{e:enmontensor}
 T_{\textrm{EM}}^{\mu\nu}
=-\frac{1}{4\pi}[SF^\mu_{\textrm{ }\textrm{ }\al}F^{\nu\al}+
P\widetilde{F}^\mu_{\textrm{ }\textrm{
}\al}F^{\nu\al}-g^{\mu\nu}\mathcal{L}_{\textrm{EM}}],
\end{equation}
for which we have
\begin{equation}\label{e:divtensor}
 \nabla_\mu T_{\textrm{EM}}^{\mu\nu}=J_\mu F^{\mu\nu}.
\end{equation}

We are considering magnetospheres described by the BI theory in the
force-free limit, i.e., the EM energy momentum tensor be conserved:
\begin{equation}\label{e:ffcond}
 J_\mu F^{\mu\nu}=0.
\end{equation}
This implies $p=0$ so that
$\mathcal{L}_{\textrm{EM}}(s,p)=\mathcal{L}_{\textrm{EM}}(s)$.

\section{Steady magnetospheres on rotating black holes}
\label{sec:BHmagnetosphere}

Let us consider the steady and axisymmetric magnetosphere on a Kerr
black hole, whose metric on the BL coordinates is
\begin{equation}\label{e:Kerr}
 ds^2=-\frac{\rho^2\triangle}{A}dt^2+\frac{\rho^2}{\triangle}dr^2+
\rho^2 d\th^2+\frac{A\sin^2\th}{\rho^2}(d\phi-\om dt)^2,
\end{equation}
where $\rho^2=r^2+a^2\cos^2\th$, $\triangle=r^2-2Mr+a^2$,
$A=2Mr(r^2+a^2)+\rho^2\triangle$ and $\om=2Mra/A$.

\subsection{The stream equation}

On the Kerr black hole, the force-free condition (\ref{e:ffcond})
reads:
\begin{equation}\label{e:ffcond1}
 \pa_r A_0J^r+\pa_\th A_0 J^\th=0,
\end{equation}
\begin{equation}\label{e:ffcond2}
 \pa_r A_0J^0+F_{r\th}J^\th+\pa_r A_\phi J^\phi=0,
\end{equation}
\begin{equation}\label{e:ffcond3}
 \pa_\th A_0J^0-F_{r\th}J^r+\pa_\th A_\phi J^\phi=0,
\end{equation}
\begin{equation}\label{e:ffcond4}
 \pa_r A_\phi J^r+\pa_\th A_\phi J^\th=0.
\end{equation}

It is convenient to use the Poison bracket defined by
\begin{equation}
 [C,D]\equiv\pa_r C\pa_\th D-\pa_\th C\pa_r D.
\end{equation}
When $C$ is a function of $D$, we must have $[C,D]=0$. From Eqs.\
(\ref{e:ffcond1}) and (\ref{e:ffcond4}), we get
\begin{equation}
 [A_0, A_\phi]=0.
\end{equation}
So $A_0$ should be a function of $A_\phi$. We can define:
\begin{equation}
 dA_0=-\Om(A_\phi)dA_\phi,
\end{equation}
where $\Om$ is the angular velocity of a magnetic field line, which
is constant along any field line.

Eq.\ (\ref{e:currents}) can be expressed as
\begin{equation}\label{e:currents1}
 J^0=\frac{1}{4\pi}\nabla\cdot\left[Sg^{00}(\om-\Om)
\nabla A_\phi\right],
\end{equation}
\begin{equation}\label{e:currents2}
 J^r=-\frac{1}{4\pi\sqrt{-g}}\pa_\th(SB_T),
\end{equation}
\begin{equation}\label{e:currents3}
 J^\th=\frac{1}{4\pi\sqrt{-g}}\pa_r(SB_T),
\end{equation}
\begin{equation}\label{e:currents4}
 J^\phi=\frac{1}{4\pi}\nabla\cdot\left[S\left(g^{\phi\phi}-g^{0\phi}
\Om\right)\nabla A_\phi\right],
\end{equation}
where the operator $\nabla_i=(\nabla_r, \nabla_\th)$ is associated
with the full Kerr metric. The toroidal field $B_T=(\triangle
\sin\th/\rho^2)F_{r\th}$.

From Eqs.\ (\ref{e:ffcond1}), (\ref{e:ffcond4}), (\ref{e:currents2})
and (\ref{e:currents3}), we find that
\begin{equation}
 [A_\phi,\textrm{}SB_T]=0.
\end{equation}
So $\sin\th SF_{r\th}$ is also a function of $A_\phi$. Let us denote
\begin{equation}
 \psi\equiv2\pi A_\phi, \textrm{ }\textrm{ }\textrm{ }
I(\psi)\equiv-2\pi SB_T.
\end{equation}
From Eq.\ (\ref{e:ffcond2}) or (\ref{e:ffcond3}), we can have
\begin{equation}\label{e:jphi}
 J^\phi=\Om J^0-\frac{II'}{8\pi^2S\triangle\sin^2\th},
\end{equation}
where the prime denotes the derivative with respect to $\psi$.

By comparing Eqs.\ (\ref{e:currents4}) and (\ref{e:jphi}) after
insertion of Eq.\ (\ref{e:currents1}), we derive the stream equation
of the black hole magnetosphere in the non-linear theory:
\begin{equation}\label{e:GSeq}
 S\nabla\cdot\left\{\frac{\rho^2S}{A\sin^2\th}\left[1-
\frac{A^2\sin^2\th(\Om-\om)^2}{\rho^4\triangle}\right]\nabla\psi
\right\}+\frac{AS^2(\Om-\om)}{\rho^2\triangle}\Om'(\nabla\psi)^2+
\frac{II'}{\triangle\sin^2\th}=0.
\end{equation}
The equation is just modified with the factor $S$. When
$S\rightarrow-1$, the equation recovers the case in the Maxwell
theory. As shown by the equation, the positions of the lightsurfaces
are not changed.

With the above equations, we get
\begin{equation}\label{e:s}
 s=\frac{1}{8\pi^2A\sin^2\th}\left\{\frac{AI^2}{\triangle S^2}+
\left[1-\frac{A^2\sin^2\th}{\rho^4\triangle}(\Om-\om)^2\right]
[\triangle(\pa_r\psi)^2+(\pa_\th\psi)^2]\right\}.
\end{equation}
From this, the expression of $S$ for the BI theory is obtained:
\begin{equation}\label{e:BIS2}
 S^2=\frac{A(f\triangle-I^2)}{Af\triangle+
\left[\triangle-\frac{A^2\sin^2\th}{\rho^4}(\Om-\om)^2\right]
[\triangle(\pa_r\psi)^2+(\pa_\th\psi)^2]},
\end{equation}
where $f(\th)=4\pi^2b^2\sin^2\th$.

\subsection{The energy and momentum extraction rates}

The field components observed by the Zero Angular Momentum Observers
(ZAMOs) in the unit basis vectors of the absolute space
\cite{MacDonald:1982zz} are
\begin{equation}\label{e:E}
 \mathbf{E}=-\frac{\mathbf{D}}{S}=-\frac{\Om-\om}{2\pi\La\sqrt{\rho^2}}
\left(\sqrt{\triangle}\pa_r\psi\mathbf{e}_r+\pa_\th\psi\mathbf{e}_\th\right),
\end{equation}
\begin{equation}\label{e:B}
 \mathbf{B}=-\frac{\mathbf{H}}{S}=\frac{1}{2\pi\sqrt{A}\sin\th}\left(
\pa_\th\psi\mathbf{e}_r-\sqrt{\triangle}\pa_r\psi\mathbf{e}_\th
-\frac{I\sqrt{\rho^2}}{S\La}\mathbf{e}_\phi\right),
\end{equation}
where $\La^2=\rho^2\triangle/A$.

From the energy-momentum tensor (\ref{e:enmontensor}), we can obtain
the poloidal components of the energy and angular momentum flux
densities:
\begin{equation}\label{e:r-enmom}
 \mathcal{E}^r=\Om\mathcal{L}^r=-\frac{\Om I}
{16\pi^2\rho^2\sin\th}\pa_\th\psi,
\end{equation}
\begin{equation}\label{e:th-enmom}
 \mathcal{E}^\th=\Om\mathcal{L}^\th=\frac{\Om I}
{16\pi^2\rho^2\sin\th}\pa_r\psi.
\end{equation}
The total rate of angular momentum and energy extraction from the
hole is given by the integration of the radial densities over all
the accessable spacetime:
\begin{equation}\label{e:totr-enmom}
 L=-\frac{1}{8\pi^2}\int Id\psi, \textrm{ }\textrm{ }\textrm{ }
E=\frac{1}{4\pi}\int (\mathbf{E}\times\mathbf{H})\cdot d\mathbf{s}
=-\frac{1}{8\pi^2}\int\Om Id\ps.
\end{equation}
So they take the same form in appearance as in the Maxwell theory,
containing no $S$. But, it should be noticed that the functions
actually have been corrected by the non-linear factor $S$.

\section{Boundary behaviors}
\label{sec:bc}

As done in the Maxwell theory in the previous work
\cite{Li:2017qzu}, the conditions of the differential equation
(\ref{e:GSeq}) at the event horizon and spatial infinity can be
determined. In what follows, we examine the conditions in the BI
theory.

\subsection{The condition at the event horizon}

\subsubsection{The Znajek condition}

The equation (\ref{e:GSeq}) at the event horizon $r\rightarrow
r_+=M+\sqrt{M^2-a^2}$
is simplified to
\begin{equation}\label{e:horeq}
 II'
=\frac{S\sqrt{A}\sin\th(\Om-\om)}
{\rho^2}\pa_\th\frac{S\sqrt{A}\sin\th(\Om-\om)\pa_\th\psi}{\rho^2}.
\end{equation}
This exactly gives the Znajek regularity condition
\cite{Znajek:1977unknown} at the horizon:
\begin{equation}\label{e:Zbd}
 I_+=-\frac{2Mr_+S_+\sin\th(\Om_+-\om_+)}{\rho_+^2}\pa_\th\psi_+,
\end{equation}
where the quantities with the subscript $+$ indices denote values at
the $r=r_+$. Note that $S$ is negative here. It is interesting to
find that the expression of $S_+$ obtained from Eq.\ (\ref{e:BIS2})
with $r\rightarrow r_+$ gives the same condition. As it is seen, the
Znajek condition is modified in the non-linear theory compared to
the Maxwell theory case.

This relation is actually the result that the electromagnetic fields
satisfy the radiation condition \cite{Nathanail:2014aua} at the
horizon:
\begin{equation}\label{e:radcon}
 E_\th=\pm B_\phi.
\end{equation}
As determined in the ingoing frame \cite{Znajek:1977unknown}, the
field components (\ref{e:E}) and (\ref{e:B}) at the horizon
generally satisfy the following conditions \cite{MacDonald:1982zz}:
\begin{equation}\label{e:horcon1}
 E_r, B_r, E_H, B_H\sim\mathcal{O}(1),
\end{equation}
\begin{equation}\label{e:horcon2}
  \mathbf{B}_H=\mathbf{E}_H\times\mathbf{e}_r,
\end{equation}
where the horizon fields are defined by
\begin{equation}
  \mathbf{E}_H=\La E_\th\mathbf{e}_\th|_{r\rightarrow r_+},
\textrm{ }\textrm{ }\textrm{ } \mathbf{B}_H=\La
B_\phi\mathbf{e}_\phi|_{r\rightarrow r_+}.
\end{equation}
So the condition (\ref{e:horcon2}) corresponds to the negative sign
case of the radiation condition (\ref{e:radcon}), which also leads
to the Znajek condition (\ref{e:Zbd}). Similarly, we can have the
field components $D_r$, $H_r$, $\mathbf{D}_H$ and $\mathbf{H}_H$
according to the relations given in Eqs.\ (\ref{e:E}) and
(\ref{e:B}).

\subsubsection{The horizon resistivity}

The boundary conditions of the electrodynamics on the horizon give
rise to the notions of surface charge and current
\cite{1978MNRAS.185..833Z,Damour:1978cg}. In the non-linear theory,
their definitions are changed to
\begin{equation}
  \si_H=\frac{D_r}{4\pi},
\end{equation}
\begin{equation}
  \mathbf{j}_H=-\frac{1}{4\pi}\mathbf{H}_H\times\mathbf{e}_r.
\end{equation}
When $\psi=\psi(\th)$ at the horizon, $D_r=0$ and so the surface
charge is zero. It can be checked that the surface charge and
current satisfy the charge conservation equation
\cite{MacDonald:1982zz}. Combined with equation (\ref{e:horcon2}),
we have
\begin{equation}
  \mathbf{j}_H=\frac{\mathbf{E}_H}{R_H},
\end{equation}
where the resistivity of the horizon is now
\begin{equation}\label{e:res}
  R_H=-\frac{4\pi}{S_+}.
\end{equation}
So the resistivity is not constant any more on the horizon. As
indicated by the monopole solution that will be derived in the next
section, $-S_+>1$ in the BI theory. So the resistivity here should
be larger than in the Maxwell theory, for which $S_+=-1$ and the
resistivity gets the minimum value $R_H=4\pi\simeq377$ ohms.

The result is different from that in \cite{Kim:2000yd}, where the
obtained resistivity is the same as in the Maxwell theory. The
reason is that the authors have chosen a special frame in which the
BI theory looks like the Maxwell theory. This should be not true for
a general observer's frame.

\subsection{The condition at spatial infinity}

Similar to the Maxwell theory case \cite{Menon:2005va,Menon:2005mg},
the finiteness of both the energy and the momentum fluxes in Eqs.\
(\ref{e:r-enmom}) and (\ref{e:th-enmom}) requires $\Om$ be
independent of $r$ at infinity:
\begin{equation}\label{e:Om-infcon}
 \Om(r,\th)\rightarrow\Om_0(\th) \textrm{ }\textrm{ } \textrm{ as }
\textrm{ }\textrm{ } r\rightarrow\infty,
\end{equation}
where $\Om_0(\th)$ is the value at the infinite boundary. Since
$I(\Om)$ and $\psi(\Om)$ are functions of $\Om$, then the values
$I_0$ and $\psi_0$ at infinity should be independent of $r$ as well.
In combination with the expression (\ref{e:BIS2}) of $S$, we have
\begin{equation}\label{e:psi-I-infcon}
 r\rightarrow\infty: \textrm{ }\textrm{ }\textrm{ }
\psi(\Om)\rightarrow\psi_0(\Om_0(\th)), \textrm{ }\textrm{ }\textrm{
} I(\Om)\rightarrow I_0(\Om_0(\th)), \textrm{ }\textrm{ }\textrm{ }
S^2\rightarrow1.
\end{equation}
At spatial infinity, the BI theory with weak fields approaches the
Maxwell theory.

In this case, the stream equation (\ref{e:GSeq}) at infinity reduces
to
\begin{equation}\label{e:}
 I_0\frac{\pa I_0}{\pa\psi_0}
=\sin\th\Om_0\pa_\th(\sin\th\Om_0\pa_\th\psi_0).
\end{equation}
Similarly, the equation gives rise to the relation:
\begin{equation}\label{e:infbd}
 I_0=-\sin\th\Om_0\pa_\th\psi_0.
\end{equation}
Here, the negative sign is chosen for $\Om_+\leq\om_+$. It
corresponds to the positive sign case of the radiation condition
(\ref{e:radcon}), which guarantees outflow of energy from the hole.

\subsection{Matching the boundary conditions}
\label{sec:}

It is seen that the two boundaries of any field line in the black
hole magnetosphere are in two different regimes: one is in the
Maxwell theory and the other is in the BI theory.

As in \cite{Li:2017qzu}, we first consider the case that the
functions at the horizon and at infinity are matched to be
identical:
\begin{equation}\label{e:con1}
 \psi_0=\psi_+, \textrm{ }\textrm{ }\textrm{ }
\Om_0=\Om_+, \textrm{ }\textrm{ }\textrm{ } I_0=I_+.
\end{equation}
Then, from Eqs.\ (\ref{e:Zbd}) and (\ref{e:infbd}), we obtain
\begin{equation}\label{e:}
 S_+=-\frac{\Om_+(1-a\sin^2\th\om_+)}{\om_+-\Om_+}.
\end{equation}
Thus, for a given black hole, the quantum correction to the
magnetosphere near the horizon is relying on the angular velocity
$\Om_+$ of the field lines. The correction factor $-S_+$ gets larger
when $\Om_+$ increases.

Instead, we may also make the identifications:
\begin{equation}\label{e:con2}
 \Om_0=\Om_+, \textrm{ }\textrm{ }\textrm{ }
\frac{\pa_\th\psi_0}{I_0}=S_+\frac{\pa_\th\psi_+}{I_+}.
\end{equation}
By comparing Eqs.\ (\ref{e:Zbd}) and (\ref{e:infbd}), we then have
the solution
\begin{equation}\label{e:OmP}
 \Om_+=\Om_0
=\frac{a}{2Mr_++\rho^2_+}.
\end{equation}
If we choose positive sign in Eq.\ (\ref{e:infbd}), the resulting
solution is
\begin{equation}\label{e:OmN}
 \Om_+=\Om_0
=\frac{1}{a\sin^2\th}.
\end{equation}
In this case, the angular velocity is larger than the one of the
black hole. These two solutions at the boundaries are exactly the
asymptotical solutions found in \cite{Menon:2005va,Menon:2005mg}. It
is easy to check that the latter solution with $S=-1$ is still an
exact solution to the stream equation in all regions in the BI
theory. But this trivial solution is unphysical since it admits null
current.

\section{The perturbative monopole solution in the BI theory}
\label{sec:monopolesol}

It is easy to find that the monopole solution $\psi=-\cos\th$ still
exists to the stream equation (\ref{e:GSeq}) on Schwarzschild back
holes with $a=0$ (and so $\Om=I=0$). Based on the solution, the
perturbative monopole solution (though its existence is debated
\cite{Grignani:2018ntq}) can be derived in slowly rotating black
holes, as done in \cite{Blandford:1977ds}. In this section, we shall
explore the corresponding monopole solution in the BI theory.

Let us define the dimensionless parameters:
\begin{equation}\label{e:}
 x=\frac{r}{r_0}, \textrm{ }\textrm{ }\textrm{ }
\al=\frac{a}{r_0},
\end{equation}
where $r_0=2M$ is the radius of the horizon in the Schwarzschild
case. The functions can be expanded in powers of $\al$:
\begin{equation}\label{e:}
 \psi=\psi_0+\al^2\psi_2+\cdots,
\end{equation}
\begin{equation}\label{e:}
 \widetilde{\Om}= r_0\Om=\al\widetilde{\Om}_1+\al^3\widetilde{\Om}_3
+\cdots,
\end{equation}
\begin{equation}\label{e:}
 \widetilde{I}= r_0I=\al\widetilde{I}_1+\al^3\widetilde{I}_3
+\cdots,
\end{equation}
where $\psi_0$ is the monopole solution of the zero-th order
equation:
\begin{equation}\label{e:}
 \psi_0=-\cos\th.
\end{equation}

With them, we get the expanded form of $S^2$:
\begin{equation}\label{e:expBIS2}
 S^2=\frac{kx^4}{1+kx^4}+\al^2\frac{g}{(x-1)(1+kx^4)^2}+\cdots,
\end{equation}
where
\begin{equation}\label{e:k}
 k=4\pi^2b^2r_0^4,
\end{equation}
\begin{eqnarray}\label{e:g}
 g=kx(x-1)[2x-(x-1)\sin^2\th]+kx\sin^2\th(1-\widetilde{\Om}_1x^3)^2
\nonumber \\
-2kx^4(x-1)\frac{\pa_\th\psi_2}{\sin\th}-x^3(1+kx^4)
\frac{\widetilde{I}_1^2}{\sin^2\th}.
\end{eqnarray}

The expanded equation (\ref{e:GSeq}) with (\ref{e:expBIS2}) at the
order $\mathcal{O}(\al^2)$ gives the equation:
\begin{equation}\label{e:pereq}
 L^2\psi_2=\frac{1}{1-\frac{1}{x}}\left[\sin^2\th\left(\widetilde{\Om}_1
-\frac{1}{x^3}\right)\pa_\th\widetilde{\Om}_1+\sin2\th\left(\widetilde{\Om}_1
-\frac{1}{x^3}\right)^2-\frac{\widetilde{I}_1\widetilde{I}_1'}{\sin\th}\right]
+\frac{\sin2\th}{x^5}+\de,
\end{equation}
where
\begin{equation}\label{e:}
 L^2=\frac{1}{\sin\th}\pa_x\left(1-\frac{1}{x}\right)\pa_x
+\frac{1}{x^2}\pa_\th\left(\frac{1}{\sin\th}\pa_\th\right),
\end{equation}
and the correction terms
\begin{equation}\label{e:}
 \de=-\frac{1}{kx^4}\left[\frac{1}{1-\frac{1}{x}}\left(
\frac{\widetilde{I}_1\widetilde{I}_1'}{\sin\th}+\frac{\pa_\th
g}{2x^3(1+kx^4)}\right)+ \frac{2kx^3(1-\frac{1}{x})}
{\sin\th(1+kx^4)}\pa_x\psi_2\right].
\end{equation}

Towards the horizon with $x\rightarrow1$, the equation diverges as
$\mathcal{O}(1/(1-1/x))$. So, to avoid divergence, the relevant
terms must cancel out, i.e.,
\begin{equation}\label{e:perZbd}
 \widetilde{I}_1(x=1,\th)=\sqrt{\frac{k}{1+k}}
(\widetilde{\Om}_1-1)\sin\th\pa_\th\psi_0.
\end{equation}
This is also the boundary condition (\ref{e:Zbd}) at the horizon.

Since the magnetosphere in asymptotical regions is in the regime of
the Maxwell theory, we can still take the outer boundary condition
as the Michel monopole solution obtained in the flat spacetime.
Inserting it into the condition (\ref{e:infbd}) at infinity leads to
\begin{equation}\label{e:perinfbd}
 \widetilde{I}_1(x\rightarrow\infty,\th)=-\widetilde{\Om}_1\sin^2\th.
\end{equation}

Adopting the matching condition (\ref{e:con1}) given in the previous
section for the above two boundary conditions (\ref{e:perZbd}) and
(\ref{e:perinfbd}), we derive
\begin{equation}\label{e:}
 \widetilde{\Om}_1=\frac{\sqrt{k}}{\sqrt{1+k}+\sqrt{k}}.
\end{equation}
So the angular velocity grows with the value of $k$ and reaches the
maximum value $1/2$ as $k\rightarrow\infty$. With the angular
velocity, we have
\begin{equation}\label{e:}
 \widetilde{I}_1=-\frac{\sqrt{k}}{\sqrt{1+k}+\sqrt{k}}\sin^2\th.
\end{equation}

Inserting the above results into Eq.\ (\ref{e:pereq}), the solution
of $\psi_2$ can be derived basically. But it is hard to do so
because the resulting equation is highly non-linear. Here, we only
consider the solution in the asymptotical region. At large $x$, the
equation reduces to
\begin{equation}\label{e:}
 \frac{1}{\sin\th}\pa_x^2\psi_2+\frac{1}{x^2}\pa_\th
\left(\frac{1}{\sin\th}\pa_\th\psi_2\right)=
-\frac{2\widetilde{\Om}_1\sin2\th}{x^3}.
\end{equation}
The solution is
\begin{equation}\label{e:}
 \psi_2=\frac{\sqrt{k}\sin^2\th\cos\th}{(\sqrt{1+k}+\sqrt{k})x}.
\end{equation}
When $k\rightarrow\infty$, this recovers the result in
\cite{Blandford:1977ds}.

\section{Discussion and conclusion}
\label{sec:conclusion}

Force-free non-linear electrodynamics on rotating black holes is
discussed. Based on the derived stream equation, we analyze the
boundary conditions at the horizon and at spatial infinity. Compared
to the case in the Maxwell theory, the Znajek condition at the
horizon is modified in the non-linear theory, while the one at
infinity remains the same as in the Maxwell theory. We also show
that the surface resistivity on the horizon is modified in the ZAMO
frame.

In terms of the boundary conditions, we further obtain the
perturbative solution of the split monopole in the slow rotation
limit. With the solution, we can find that the horizon resistivity
given by Eq.\ (\ref{e:res}) is larger than that in the Maxwell
theory: $R_H=4\pi\sqrt{k}/\sqrt{1+k}$. Following the analysis in
\cite{1978MNRAS.185..833Z}, this implies that the energy extraction
rate should be lower in the BI theory. For given electric potential
difference induced by the hole rotation, the energy output should be
smaller with a larger impedance. Indeed, from Eq.\
(\ref{e:totr-enmom}) with the monopole solution, we can see that the
energy extraction rate is smaller compared to that in the Maxwell
theory:
\begin{equation}\label{e:}
 \frac{E^\textrm{(BI)}}{E^\textrm{(Maxwell)}}=
\left(\frac{2\sqrt{k}}{\sqrt{1+k}+\sqrt{k}}\right)^3.
\end{equation}
The ratio only relies on the parameter $k$ defined in Eq.\
(\ref{e:k}). It gets the Maximum value as $k\rightarrow\infty$. It
is interesting that the parameter $k$ is related to the parameter
$b$ in the BI theory as well as the radius $r_0$ of the black hole
horizon, which implies that we can recover the results in the
Maxwell theory only with  $k\rightarrow\infty$, without need of a
large $b$. The non-linear correction in the BI theory become more
important on a lighter black hole (with smaller $r_0$). On the
opposite, the BI theory with finite $b$ can behave like the Maxwell
theory on a massive black hole. This is different from the situation
in the Minkowski spacetime.

The reason for this difference might be due to the converging effect
of the horizon on the field lines. For given boundary conditions at
infinity, the field lines are much denser across a horizon on a
lighter black hole because the area of the horizon is smaller. This
makes the fields to be stronger and easier to reach the QED regime.
Similarly, for a massive black hole, the field field lines will be
diluted on the horizon with a large area.

Finally, it should be pointed out that the result is also consistent
with that from the quantum aspects of black holes. Thermal particles
excited in the vacuum near the horizon should also contribute to the
quantum corrections to the electrodynamics in this region. So the
quantum corrections become more important for the electrodynamics on
a smaller black hole who has a larger Hawking temperature. Of
course, our discussion here does not include the quantum effects
from the black hole spacetime. It is interesting to do further
investigation in future study.

\section*{Acknowledgements\markboth{Acknowledgements}{Acknowledgements}}

This work is supported by the Yunnan Natural Science Foundation
2017FB005 and 2014FB188.

\bibliographystyle{JHEP}
\bibliography{b}

\end{document}